\documentstyle[aps,multicol,epsf]{revtex}

\newcommand{\be}{\begin{equation}}
\newcommand{\ee}{\end{equation}}
\newcommand{\bea}{\begin{eqnarray}}
\newcommand{\eea}{\end{eqnarray}}

\def\cR{{\cal R}}
\def\cE{{\cal E}}

\begin{document}
\title{
  \begin{flushright} \begin{small}
   LGCR--99/08/02\\  DTP--MSU/99-22 \\ hep-th/9910171
  \end{small} \end{flushright}
\vspace{1.cm}
Classical glueballs in non-Abelian Born-Infeld theory
}
\author{
        Dmitri Gal'tsov$^{a,b}$
    and Richard Kerner$^{a}$
}
\address{$^{a}$Laboratoire de  Gravitation and Cosmologie Relativistes,\\
Universit\`e Pierre et Marie Curie, 4, place Jussieu,
Paris, 75252, France}
\address{$^{b}$Department of Theoretical Physics,
         Moscow State University, 119899, Moscow, Russia,}

\maketitle

\begin{abstract}
It is shown that the Born--Infeld--type modification of the quadratic
Yang--Mills action suggested by the superstring theory gives rise to
classical particle-like solutions prohibited in the standard Yang--Mills
theory. This becomes possible due to the scale invariance breaking by
the Born--Infeld non--linearity. New classical glueballs are sphaleronic
in nature and exhibit a striking similarity with the Bartnik-McKinnon
solutions of the Yang--Mills theory coupled to gravity.

\end{abstract}

\begin{multicols}{2}
\narrowtext
\section{Introduction}

   The standard Yang--Mills theory does not admit classical
particle-like solutions \cite{De76,Pa77,Co77}. More precisely, this famous
no-go result asserts that there exist no finite--energy non--singular
solutions to the four--dimensional Yang--Mills equations which would be
either static, or non-radiating time--dependent. \cite{Co77}.
Non-existence of static solutions can be related to conformal invariance
of the Yang--Mills theory, which implies that the stress--energy tensor is
traceless :
$T_\mu^\mu=0=-T_{00}+ T_{ii}$,
where $\mu=0,...,3,\; i=1,2,3$. Given the positivity of the energy density
$T_{00}$, this means that the sum of the principal pressures $T_{ii}$
is everywhere positive, {\it i.e.} the Yang--Mills matter is repulsive.
This makes the mechanical equilibrium impossible \cite{Gi82}.

The Higgs field breaks the conformal invariance of the pure Yang--Mills
theory and so in the spontaneously broken gauge theories particle-like
solutions may exist. Two types of such solutions are known: magnetic monopoles
and sphalerons. Topological criterion for the existence of monopoles
is the non-triviality of the second homotopy group of the broken phase
manifold $\pi_2(G/H)$ associated with the configuration of the Higgs field.
Thus topologically stable monopoles exist in the $SO(3)$ gauge theory with
a real Higgs triplet, in which case $G/H=S^2$, but do not exist in the $SU(2)$
gauge theory with a complex Higgs doublet, where the symmetry is completely
broken (the Higgs broken phase manifold is $S^3$).

However, in the theory with doublet Higgs another particle--like solution
has been found by Dashen, Hasslacher and Neveu \cite{DaHaNe74}. Its existence
was explained by Manton \cite{Ma83} as a consequence of non--triviality
of the {\em third} homotopy group $\pi_3 (S^3)$, indicating the presence of
non--contractible loops in the configuration space. This solution is {\it the}
sphaleron; it sits at the top of the potential barrier separating
topologically distinct Yang--Mills vacua. Because of this position,
the sphaleron is necessarily unstable. Still its r\^ole is very important,
since in presence of fermions it can mediate transitions without the
conservation of fermion number.

In the latter case, the  manifold of the Higgs broken phase coincides with
the gauge group manifold, and it is not quite clear, whether it is the
topology of the {\em Higgs} field, or the topology of the {\em Yang--Mills}
field itself which is crucial for the existence of this solution. This issue
was clarified after the discovery of sphaleron--like solutions in the
$SU(2)$ gauge theory coupled to gravity, without Higgs fields at all.
Particle--like solutions in this theory were found numerically by Bartnik
and McKinnon (BK) \cite{BaMc88}; their relation to sphalerons has been
explained by Gal'tsov and Volkov \cite{GaVo91} and Sudarsky and Wald
\cite{SuWa92} (for a recent review, see \cite{VoGa98}). This and other
examples (similar solutions exist in the flat space Yang--Mills theory
coupled to the dilaton) show that the topological reason for the existence
of sphalerons in the theories with gauge fields
is the non-triviality of third homotopy class of the
Yang--Mills gauge group (note that $\pi_3(G)=Z$ for any simple compact
Lie group $G$). The Higgs field in this case just plays a r\^ole of
attractive agent balancing the repulsive Yang--Mills forces. In other words,
its function is to break the {\em scale} invariance of the Yang--Mills theory
rather than the gauge invariance. The same symmetry breaking may occur due
to gravity or the presence of dilaton field, which do not imply a spontaneous
breaking of the gauge symmetry.

The superstring theory gives rise to one important modification of the
standard Yang-Mills quadratic Lagrangian suggesting the action of the
Born-Infeld (BI) type \cite{Ts97,GaGoTo98,BrPe98}. Such a
modification also breaks the scale invariance, so the natural question
arises whether in the Born--Infeld--Yang--Mills (BIYM) theory the
non-existence of classical particle-like solutions can be overruled.
This is particularly intriguing since now neither gravity, nor scalar
fields are involved, so one is thinking about the genuine {\em classical
glueballs}. Note that a mere scale invariance breaking, being a necessary
condition, by no means guarantees the existence of particle-like solutions,
and a more detailed study is needed to prove or disprove this conjecture.
Our investigation shows that the $SU(2)$ BIYM classical glueballs indeed do
exist and display a remarkable similarity with the BK solutions of the
Einstein--Yang--Mills (EYM) theory.

Non--Abelian generalisation of the Born--Infeld action presents an ambiguity
in specifying how the trace over the the matrix--valued fields is performed
in order to define the Lagrangian. Here we adopt the version with the ordinary
trace which leads to a simple closed form for the action. In fact, another
trace prescription is favored in the superstring context, namely, the
{\em symmetrized} trace \cite{Ts97}, but so far the explicit Lagrangian with
such trace is known only as perturbative series \cite{GrMoSc99}. For our
purposes the full non-perturbative Lagrangian is needed, so we consider the
ordinary trace, presenting some arguments at the end of the paper about the
possibility of extension of our results to the theory with symmetrized trace.

The BIYM action with the ordinary trace looks like a straightforward
generalisation of the corresponding $U(1)$ action in the ``square root'' form
\be \label{S}
S=\frac{\beta^2}{4\pi}\int\;(1-\cR)\;d^4x,
\ee
where
\be
\cR=\sqrt{1+\frac{1}{2\beta^2}
F^a_{\mu\nu}F_a^{\mu\nu}
-\frac{1}{16\beta^4}(F^a_{\mu\nu}{\tilde F}_a^{\mu\nu})^2}.
\ee
Here the dimensionless gauge coupling constant (in units $\hbar=c=1$)
is set to unity, so the  only parameter of the theory is the constant
$\beta$ of dimension $L^{-2}$, the ``critical'' field strength.
It is easy to see that the BI non-linearity breaks the conformal
symmetry ensuring the non-zero trace of the stress--energy tensor
\be
T^\mu_\mu=\cR^{-1}\left[4\beta^2(1-\cR)-
F^a_{\mu\nu}F_a^{\mu\nu} \right] \neq 0.
\ee
\indent
This quantity vanishes in the limit $\beta\to 0$ when the theory reduces to
the standard one.
\newline
\noindent
For the YM field we assume the usual monopole ansatz
\be    \label{Aa}
A_0^a=0, \quad
A_i^a=\epsilon_{aik}{n^k\over r}(1-w(r)) ,
\ee
where  $n^k=x^k/r,\; r=(x^2+y^2+z^2)^{1/2}$, and $w(r)$ is the real-valued
function. After the integration over the sphere in (\ref{S}) one obtains
a two-dimensional action from which  $\beta$ can be eliminated by the
coordinate rescaling $\sqrt{\beta} t\to t,\; \sqrt{\beta} r\to r$.
As a result we find the following static action:
\be
S=\int L dr, \quad L=r^2(1-\cR),
\ee
with
\be \label{R}
\cR=\sqrt{1+ 2\frac{w'^2}{r^2} + \frac{(1-w^2)^2}{r^4}},
\ee
where prime denotes the derivative with respect to r.
The corresponding equation of motion reads
\be \label{eqm}
\left(\frac{w'}{\cR}\right)'= \frac{w(w^2-1)}{r^2\cR}.
\ee

A trivial solution to the Eq.(\ref{eqm}) $w\equiv 0$ corresponds to the
pointlike magnetic BI-monopole with the unit magnetic charge (embedded
$U(1)$ solution). In the Born--Infeld theory it has a finite self-energy.
For time-independent configurations the energy density is equal to minus the
Lagrangian, so the total energy (mass) is given by the integral
\be \label{M}
M=\int_0^\infty (\cR-1)r^2 dr.
\ee
For $w\equiv 0$ one finds
\bea \label{mu1}
M&=&\int \left(\sqrt{r^2+1}-r^2\right)dr\nonumber\\
&=&\frac{\pi^{3/2}}{3{\Gamma (3/4)}^2}\approx 1.23604978.
\eea

Let us look now for essentially non--Abelian solutions of finite mass. In
order to assure the convergence of the integral (\ref{M}) the quantity
$\cR-1$ must fall down faster than $r^{-3}$ as $r\to \infty$. Thus, far from
the core the BI corrections have to vanish and the Eq.(\ref{eqm})
should reduce to the ordinary YM equation. The latter is equivalent to the
following two-dimensional autonomous system
\cite{Chern78,Kerner,Pr79,BrFoMa94}:
\be \label{ds}
\dot w=u, \quad \dot u=u+(w^2-1)w,
\ee
where a dot denotes the derivative with respect to $\tau=\ln r$.
This dynamical system has three non-degenerate stationary points
$(u=0, w=0,\pm1)$, from which $u=w=0$ is a focus,
while two others $u=0,\,w=\pm 1$ are
saddle points with eigenvalues $\lambda =-1$ and $\lambda =2$.
The  separatices along the directions $\lambda =-1$ start at infinity
and after passing through the saddle points go to the focus with the
eigenvalues $\lambda=(1\pm i\sqrt{3})/2$. The function
$w(\tau)$ approaching the focus as $\tau\to\infty$ is unbounded.
Two other separatices, passing through saddle points
along the directions specified by $\lambda =2$, go to infinity in both
directions. Since there are no limiting circles, generic phase curves
go to infinity or approach the focus,
unless $w=0$ identically. All of them produce a divergent mass integral
(\ref{M}). The only trajectories remaining bound  as $\tau\to\infty$
are those which go to the saddle points along the separatrices specified
by $\lambda=-1$.

 From this reasoning one finds that {\em the only finite-energy configurations
with non-vanishing magnetic charge are the embedded U(1) BI-monopoles}.
Indeed, such solutions should have asymptotically $w=0$, which does
not correspond to bounded solutions unless $w\equiv 0$. The remaining
possibility is $w=\pm 1, \dot w=0$ asymptotically, which corresponds to
zero magnetic charge. Coming back to $r$-variable one finds from (\ref{eqm})
\be \label{was}
w=\pm 1+ \frac{c}{r} + O(r^{-2}),
\ee
where $c$ is a free parameter. This gives a convergent integral  (\ref{M})
as $r\to\infty$. Note that two values $w=\pm 1$ correspond to two neighboring
topologically distinct YM vacua.

Now consider local solutions near the origin $r=0$. For convergence of
the total energy (\ref{M}), $w$ should tend to a finite limit
as $r\to 0$. Then using the Eq.(\ref{eqm}) one finds that
the only  allowed limiting values are $w=\pm 1$ again. In view
of the symmetry of (\ref{eqm}) under reflection $w\to \pm w$,
one can take without loss of generality $w(0)=1$. Then the following
Taylor expansion can be checked to satisfy the Eq.(\ref{eqm}):
\be \label{w0}
w=1-br^2+\frac{b^2(44 b^2+3)}{10(4 b^2+1)} r^4 +O(r^6),
\ee
with $b$ being (the only) free parameter.

As $r\to 0$, the function $\cR$ tends to a finite value
\be
\cR=\cR_0+ O(r^2),  \ \ \, \ \ \,  \cR_0=1+12 b^2.
\ee
By rescaling $r^2\cR_0= {\tilde r}^2$ one can cast the Eq.(\ref{eqm}) again
into the form of the dynamical system (\ref{ds}), so by the same reasoning
the series (\ref{w0}) may be shown to correpond to the local solution
starting as $\tilde\tau\to -\infty,\, \tilde\tau=\ln {\tilde r}$ from
the saddle point $u=0,w=1$ along the separatrix $\lambda=2$. Another
bounded $w$ satisfying the dynamical system (\ref{ds}) might start at
the focal point. But then in terms of ${\tilde r}$
\be
w\sim C\sqrt{{\tilde r}}\sin\left(\frac{\sqrt{3}}{2}\ln {\tilde r}
+\alpha\right)
\ee
with $\alpha=const$, this does not satisfy the assumption $\cR\to const$,
therefore it is not a solution of the initial system (\ref{eqm}).
Thus we proved that {\em any regular solution  of the} Eq.(\ref{eqm})
{\em belongs to the one-parameter family of local solutions} (\ref{w0})
{\em near the origin}.

It follows that the global finite energy solution starting with (\ref{w0})
should meet some solution from the family  (\ref{was}) at infinity. Since
both these local solutions are non--generic, one can at best match
them for some discrete values of parameters. To complete the existence
proof one has to show that this discrete set of parameters is non-empty.
The idea of the proof is as follows. First, rewrite the Eq.(\ref{eqm}) in
the resolved form
\be \label{wnu}
\ddot w =\gamma \dot w + w (w^2-1),
\ee
where the ``negative friction coefficient'' is
\be \label{gamma}
\gamma =1+\frac{\dot\cR}{\cR}=1-\frac{\left[\dot w+w(1-w^2)\right]^2+
(1-w^2)^3}{r^4+(1-w^2)^2}.
\ee
It is easy to show that $w$ can not have local minima for $0<w<1, \, w<-1$
and can not have local maxima for $-1<w<0,\,w>1$. In view of
(\ref{was})(\ref{w0}) one finds that {\em any regular solution lies entirely
within the strip $-1<w<1$ and has at least one zero}. Once $w$ leaves
the strip, it has to diverge. The divergence occurs at some finite
$\tau=\tau_0$ with the following leading term :
\be
w\sim \pm\frac{1}{\sqrt{\tau_0-\tau}}.
\ee
The Eq.(\ref{wnu}) may be presented in the form of the ``energy
equation''
\be \label{cE}
\dot \cE=\gamma {\dot w}^2,\quad \cE=\frac12{\dot w}^2 -\frac14(1-w^2)^2.
\ee
For the ordinary quadratic Yang-Mills system  $\gamma \equiv 1$,
so the ``energy'' $\cE$ diverges soon after the solution leaves the strip
$[-1,\,1]$. However, in the present case $\gamma$ can become negative when
$\dot w$ and $w$ grow up, and this can stop further ``acceleration'' or
even reverse it. One has to show that this may happen  before $w$ leaves
the strip $[-1,\,1]$. Observe that in the Eq.(\ref{wnu}) all
terms except for $\dot \cR/\cR$ in $\gamma$ (\ref{gamma})
are invariant under rescaling $kr\to \hat r$, while the $\cR$-term changes to
\be
\cR\to \sqrt{1+\frac{k^4}{{\hat r}^4}\left[2{\dot w}^2 +(1-w^2)^2\right]}.
\ee
Thus, fixing the scale $k^2=b$, where $b$ is the free parameter of
the local solution (\ref{w0}), one finds that, for sufficiently large $b$,
the function $\gamma$ can be made negative in any desired region.
Now, if $b$ is {\em too} large, the sign of the derivative $\dot w$ will be
reversed, and $w$ will leave the strip in the positive direction.
For some precisely tuned value of $b$ the solution will remain a monotonous
function of $\tau$ reaching  the value $-1$ at infinity (Fig.1). This happens
for $b_1=12.7463$.

By a similar reasoning one can show that for another fine-tuned value
$b_2>b_1$ the integral curve $w(\tau)$ which has a minimum in the lower part
of the strip and then becomes positive will be stabilized by the friction
term in the upper half of the strip and tend to $w=1$. This solution will
have two nodes. Continuing this process we obtain the increasing sequence
of parameter values $b_n$ for which the solutions remain entirely within
the strip $[-1,\,1]$ tending asymptotically to $(-1)^n$. The lower values
$b_n$ found numerically are given in Tab.1.

\begin{figure}
\epsfxsize=7cm
\centerline{\epsffile{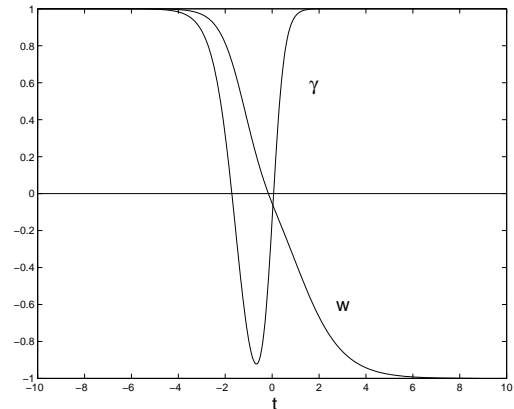}}
\caption{One-node solution monotonically interpolating between
$1$ and $-1$ and the corresponding behavior of the negative friction
coefficient $\gamma$}
\label{FIG1}
\end{figure}

\begin{center}
\begin{tabular}{|l|l|l|l|}  \hline
$n$           & $\quad b$            & $\quad M   $         \\ \hline
$1$           & $\quad 1.27463 \times 10^1\quad $ &\quad $1.13559$ \quad\\
2             & $\quad 8.87397 \times 10^2$ &  $\quad 1.21424$ \quad\\
3             & $\quad 1.87079 \times 10^4$ &  $\quad 1.23281$ \quad\\
4             & $\quad 1.27455 \times 10^6$ &  $\quad 1.23547$ \quad\\
5             & $\quad 2.65030 \times 10^7$ &  $\quad 1.23595$ \quad\\
6             & $\quad 1.80475 \times 10^9$ &  $\quad 1.23596$ \quad\\
\hline
\end{tabular}
\vglue 0.4cm
Tab 1. Parameters $b, \,M$  for first six solutions.
\end{center}

This picture displays a striking similarity with the one occuring for
the EYM system \cite{BaMc88,VoGa98}. However, there is one important
distinction. In the EYM case the sequence $b_n$ converges to a finite
value $b_\infty$, and the limiting solution exists with an infinite
number of zeros \cite{BrFoMa94}. In our case
the sequence $b_n$ has no finite limit. The region
of oscillations expands with growing $n$, and so
does the size of the particles (see Fig.~2). Typically, the first and
the last amplitude have large enough values, while in the middle zone
the amplitude of oscillations becomes very small with increasing $n$
({\it i.e.} an observer placed inside the core will see the unscreened
magnetic charge). On the contrary, with $n$ increasing the mass rapidly
converges to the finite value (\ref{mu1}) corresponding to the abelian
solution $w\equiv 0$.

\begin{figure}
\epsfxsize=7.5cm
\centerline{\epsffile{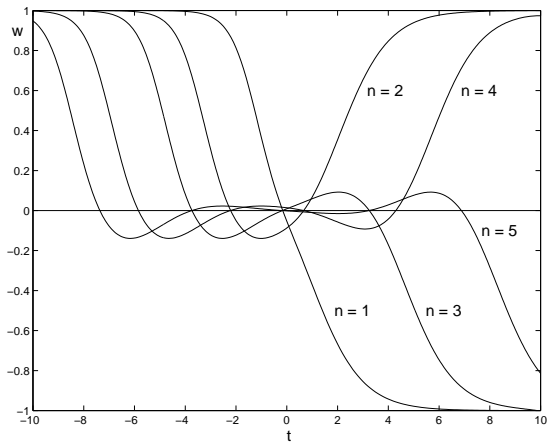}}
\caption{Solutions $w(\tau)$ for $n=1,...,5$.
}
\label{FIG2}
\end{figure}

Like in the BK case, solutions with odd and even node number $n$ have
different physical meaning \cite{VoGa98}. The lowest one with $n=1$
is the direct analog of {\em the} sphaleron. It can be shown to have the
Chern-Simons number $Q=1/2$, to possess a fermionic zero mode and it
is expected to have one odd-parity unstable decay mode along the path from
the initial to the neighboring vacuum. The potential barrier between the
neighboring vacua hence has a finite height. Higher odd-$n$ solutions
also have $Q=1/2$, but possess more than one decay direction leading to
the neighboring vacuum; they are expected to have $n$ odd-parity
negative modes. Solutions with even values of $n$ have $Q=0$, and
correspond to the paths in the phase space returning back to the same
vacuum. These may be contiuously deformed to the trivial vacuum
$w\equiv 1$ and therefore are topologically trivial.

If one uses the BIYM Lagrangian defined with the symmetrized trace, the
equation of motion still preserves the form (\ref{eqm}) with another friction
coefficient $\gamma$ and  an additional function of two variables
${\dot w}^2,\, (1-w^2)^2$ in front of the force term. It can be shown
that the minima/maxima argument used above still holds as well as
the $\gamma$-scaling argument. Therefore we expect that classical
glueballs will persist in this version of the BIYM theory too.

It can be expected that the spectrum of magnetic monopoles in the
BIYM--Higgs theory is affected by sphaleronic excitations like in the
case of gauge monopoles coupled to gravity (for discussion and
references see \cite{VoGa98}). The occurence of the limiting value of
$\beta$ found in \cite{GrMoSc99} is likely to be a typical signal.

\vskip 0.2cm
We wish to thank G.W.~Gibbons, N.S.~Manton, G.~Clement and M.S.~Volkov
for valuable comments. One of the authors (DVG) would like to thank
the Laboratory of Gravitation
and Cosmology of the University Paris-6 for hospitality and the CNRS
for support while this work was initiated.

\end{multicols}
\end{document}